\documentstyle[preprint,prb,aps]{revtex}

\begin{document}
\draft
\title{Quantum mechanics in finite dimensional Hilbert space}
\author{A. C. de la Torre, D. Goyeneche}
\address{Departamento de F\'{\i}sica,
 Universidad Nacional de Mar del Plata\\
 Funes 3350, 7600 Mar del Plata, Argentina\\
dltorre@mdp.edu.ar}
\date{\today}
\maketitle
\begin{abstract}
The quantum mechanical formalism for position and momentum of a
particle in a one dimensional cyclic lattice is constructively
developed. Some mathematical features characteristic of the finite
dimensional Hilbert space are compared with the infinite
dimensional case. The construction of an unbiased basis for state
determination is discussed.
\end{abstract}
\section{INTRODUCTION}
In the quantum mechanical description of physical systems, it is
often assumed a continuous set of states requiring an infinite
dimensional Hilbert space. There are however many physical systems
whose states belong to a discrete and finite set, with a quantum
mechanical description formalized in a finite dimensional Hilbert
space. The best known example of this is the quantum treatment of
angular momentum, indeed, a paradigm for quantum mechanics,
presented in every text book. A less known example of discrete
quantum mechanics, presented in this work, involves the
description of position and momentum observables in a finite
dimensional Hilbert space.
\par
In this case, the position observable does not take values in a
continuous set but instead it can take values on a lattice. One
important practical motivation for studying such systems is that
any computer simulation of position  and momentum will necessarily
involve a finite number of sites. On the other side, a highly
speculative motivation is that the posible existence of a
fundamental length scale, that is, a measure of length below which
the concepts of distance and localization become meaningless, can
make a discrete quantum mechanics more appropriated than a
continuous one.
 \par
In order to make this work useful from the didactic point of view,
the formalism of quantum  mechanics in a finite dimensional
Hilbert space will be presented in a constructive way where all
steps are logically connected. This work may therefore be a useful
complement to any text book where quantum mechanics in infinite
dimensional Hilbert space is developed. Finally, another didactic
feature of this work is that finite dimensional quantum mechanics
requires many interesting mathematical tools such as some finite
sums and the Discrete Fourier Transform that are not usually
presented at the undergraduate level. Furthermore, the important
differences between finite and infinite dimensional Hilbert spaces
are emphasized and the limit when the dimension becomes infinite
is considered.
\section{NOTATION AND DEFINITIONS}
In this work we will consider a particle in a one dimensional
periodic lattice with $N$ sites and lattice constant $a$. The
quantum mechanical treatment of this system requires an $N$
dimensional Hilbert space ${\cal H}$. Given any two elements of
this space $\Phi$ and $\Psi$ we will denote their internal product
by $\langle\Phi,\Psi\rangle$. We will use operators of the form
$A=\Psi\langle\Phi,\cdot\rangle$, where the dot indicates a space
holder to be occupied by the Hilbert space element upon which the
operator acts. The corresponding hermitian conjugate is
$A^{\dag}=\Phi\langle\Psi,\cdot\rangle$.
 \par
Although we don't need to choose any particular representation for
the abstract Hilbert space  ${\cal H}$, it may be convenient, for
didactic reasons, to specialize the formalism in a three or four
dimensional Hilbert space whose elements $\Psi$ are column vectors
of complex numbers. In this case $\langle\Psi,\cdot\rangle$
represents a complex conjugate row vector and operators are square
matrices. This special representation is recommended for clarity
but it is important to emphasize to the students that the
formalism of quantum mechanics can be construed in the abstract
Hilbert space and a particular representation is never required.
The mathematical beauty of quantum mechanics is, indeed, most
apparent in the abstract formulation.
\par
 Any basis $\{\varphi_{k}\}$ in the Hilbert space will have $N$ elements
labeled by an index $k$ running through the values $-j,-j+1,-j+2,
\cdots ,j-1,j,\ $ with $j=\frac{1}{2},1,\frac{3}{2},2,\frac{5}{2},
\cdots$ corresponding to the dimensions $N=2j+1= 2,3,4,5 \cdots$.
This choice of labels has some advantage and some disadvantage.
The main virtue of this symmetric labeling is that it corresponds
to the physical concepts of position and momentum that can take
positive and negative values. The main shortcoming of it, is that
there are many summations and results that are usually given in
books with integer indices running from $0$ to $N-1$. In order to
cure this deficiency, we give in Appendix A some of these sums
with the symmetric index. Furthermore, with this notation we must
keep in mind that for even $N$, $j$ takes half-odd-integer values
and this may be relevant in modular mathematics.
\par
We will adopt a very useful notation for the principal $N^{th}$
root of the identity defined by
\begin{equation}\label{ome1}
\omega =e^{i\frac{2\pi}{N}}\ .
\end{equation}
Integer powers of this quantity build a cyclic group with the
important property
\begin{equation}\label{cycl}
  1=\omega^{Nn}=\omega^{(2j+1)n}\ , \forall n=0,\pm 1,\pm 2,\cdots \ .
\end{equation}
\section{POSITION AND MOMENTUM}
The position of the particle in the lattice can take any value
(eigenvalue) $ax$ where $a$ has units of length and the discrete
number $x$ can take any value in the set $\{ -j,-j+1, \cdots
,j-1,j \}$. The state of the particle in each position is
represented by a Hilbert space element  $\varphi_{x}$ and the set
$\{\varphi_{x}\}$ is a basis in  ${\cal H}$. In the spectral
decomposition, we can write the position operator $X$ as
\begin{equation}\label{defX}
  X= \sum_{x=-j}^{j} ax
  \varphi_{x}\langle\varphi_{x},\cdot\rangle\ ,
\end{equation}
that clearly satisfies  $X \varphi_{x}= ax \varphi_{x}$. We can
now construct a \emph{translation operator} $T$ with the property
\begin{equation}\label{defT}
  T \varphi_{x}=\left\{\begin{array}{cc}
    \varphi_{x+1} &\ , x\neq j \\
    (-1)^{N-1}\varphi_{-j}=\omega^{Nj}\varphi_{-j} &\ , x=j \
  \end{array}\right. \ .
\end{equation}
We will later explain the reason for defining this operator
\emph{periodic} for odd dimension but \emph{antiperiodic} for even
dimension. This operator is given by
\begin{equation}\label{defT1}
  T =\sum_{x=-j}^{j-1}
  \varphi_{x+1}\langle\varphi_{x},\cdot\rangle +
  (-1)^{N-1}\varphi_{-j}\langle\varphi_{j},\cdot\rangle \ ,
\end{equation}
with its hermitian conjugate
\begin{equation}\label{defT3}
 T^{\dag} =\sum_{x=-j}^{j-1}
  \varphi_{x}\langle\varphi_{x+1},\cdot\rangle+
  (-1)^{N-1}\varphi_{j}\langle\varphi_{-j},\cdot\rangle \ .
\end{equation}
It is straightforward to check that this operator is unitary,
$TT^{\dag}=T^{\dag}T=\mathbf{1}$, and therefore its eigenvalues
are complex numbers of unit modulus and their eigenvectors build a
basis (see Appendix B).  Let then $\{\phi_{p}\}$ and
$\{\lambda_{p}\}$ for $p= -j,-j+1, \cdots ,j-1,j,$ be the
eigenvectors and eigenvalues of $T$. In order to determine them,
we expand  $\phi_{p}$ in terms of $\{\varphi_{x}\}$ and consider
\begin{equation}\label{xp1}
 T\sum_{x=-j}^{j}\langle\varphi_{x}, \phi_{p} \rangle\varphi_{x} =
\lambda_{p}\sum_{x=-j}^{j}\langle\varphi_{x}, \phi_{p}
\rangle\varphi_{x} \ .
\end{equation}
From Eq.(\ref{defT}) we get
\begin{equation}\label{xp2}
 \begin{array}{rl}
  \langle\varphi_{x-1}, \phi_{p} \rangle &=
  \lambda_{p}\langle\varphi_{x}, \phi_{p} \rangle\ \mbox{ for } x\neq j\ ,\\
  \langle\varphi_{j}, \phi_{p} \rangle\ \omega^{Nj} &=\lambda_{p}
  \langle\varphi_{-j}, \phi_{p} \rangle \ .
  \end{array}
\end{equation}
Up to an arbitrary phase, that can be absorbed in the definition
of $\phi_{p}$, and considering the normalization of $\phi_{p}$,
the solution of the above equations is
\begin{equation}\label{xp}
\langle\varphi_{x},\phi_{p}\rangle=\frac{1}{\sqrt{N}}\ \omega^{px}
\ ,\mbox{ and } \lambda_{p} = \omega^{-p} \ .
\end{equation}
The two bases  $\{\varphi_{x}\}$ and  $\{\phi_{p}\}$ are then
related by
\begin{equation}\label{fix1}
\varphi_{x} =\frac{1}{\sqrt{N}}\sum_{p=-j}^{j}
  \omega^{-px}\phi_{p}\ ,
\end{equation}
and
\begin{equation}\label{fip1}
\phi_{p} =\frac{1}{\sqrt{N}}\sum_{x=-j}^{j}
  \omega^{px}\varphi_{x}\ .
\end{equation}
Except for the symmetric index and a different factor, this is
essentially the Discrete Fourier Transform. Notice that if we had
not defined the translation operator antisymmetric when $N$ is
even, then we would not have obtained such a simple relation in
Eq.(\ref{xp}) above and we would have obtained different
expressions for $N$ even or odd. In other words, we choose to
define the translation operator in a way to obtain a simple
relation between the bases. The complication in the definition of
the translation operator is also related to our choice of using
symmetric indices. If we had chosen indices running from $0$ to
$N-1$, then a translation operator periodic for all $N$  would
have lead to  the two bases related by the Discrete Fourier
Transform (also expressed in terms of asymmetric indices).
\par
Therefore we have
\begin{equation}\label{eigvT}
 T\phi_{p} =\omega^{-p} \phi_{p}\ ,
\end{equation}
or equivalent,
\begin{equation}\label{defT2}
 T=\sum_{p=-j}^{j}\omega^{-p}
  \phi_{p}\langle\phi_{p},\cdot\rangle \ .
\end{equation}
We can now construct an hermitian operator $P$ as a superposition
of projectors in the basis $\{\phi_{p}\}$
\begin{equation}\label{defP}
P =\sum_{p=-j}^{j}gp\phi_{p}\langle\phi_{p},\cdot\rangle
 \ ,
\end{equation}
where $g$ is a real constant to be determined later. Clearly, this
operator is hermitian and satisfies the eigenvalue equation
$P\phi_{p}=gp\phi_{p}$. From this equation and from
Eq.(\ref{eigvT}), and doing the power expansion of the exponential
we prove that
\begin{equation}\label{TeP}
  T= \exp\left(-i\frac{2\pi}{N}\frac{P}{g}\right)\ .
\end{equation}
We can now assign a physical interpretation to the operator $P$.
This last Eq.(\ref{TeP}), together with Eq.(\ref{defT}) means that
$P$ is the \emph{generator of translations} in the position
observable. We identify this observable $P$, as is done in
classical mechanics, with the \emph{momentum}. If the position
observable takes value in a lattice with lattice constant $a$,
then, the momentum observable must also assume values in a lattice
with lattice constant $g$. In the next section we will see that
these values must be related by $ga=2\pi/N$, that is, the momentum
lattice is the reciprocal lattice  of the position.
\par
In an identical manner as was done before, we can now construct a
unitary operator $B$ that ``boosts'' the momentum states
\begin{equation}\label{defB}
  B \phi_{p}=\left\{\begin{array}{cc}
    \phi_{p+1} &\ , p\neq j \\
    (-1)^{N-1}\phi_{-j}=\omega^{Nj}\phi_{-j} &\ , p=j \
  \end{array}\right. \ ,
\end{equation}
and show that
\begin{equation}\label{eigvB}
 B\varphi_{x} =\omega^{x} \varphi_{x}\ ,
\end{equation}
and
\begin{equation}\label{defB2}
 B=\sum_{x=-j}^{j}\omega^{x}
  \varphi_{x}\langle\varphi_{x},\cdot\rangle \ ,
\end{equation}
 and also that
\begin{equation}\label{BeX}
  B=\exp\left(i\frac{2\pi}{N}\frac{X}{a}\right)\ .
\end{equation}
\section{COMMUTATION RELATION AND THE LIMIT $N\rightarrow\infty$}
Every quantum mechanics textbook emphasizes that the position and
the momentum operators satisfy the commutation relation $[X,P]=i$
(we use units such that $\hbar=1$). However, in most cases it is
not mentioned that this commutation relation is \emph{false} in a
finite dimensional Hilbert space. It becomes clear that this must
be so because one can prove that the commutation relation
$[X,P]=i$ implies that the operators $X$ and $P$ are unbound.
However, in a finite dimensional Hilbert space \emph{all}
operators are bounded; therefore such a commutation relation is
imposible in a finite dimensional Hilbert space. An explicit
calculation of the commutator in the position representation, that
is, in terms of the basis $\{ \varphi_{x}\}$ results in
\begin{equation}
[X,P] = ag\sum_{k=-j}^{j}\ \sum_{s=-j}^{j}\ \sum_{r=-j}^{j}
k(s-r)\frac{1}{N} \exp\left(i\frac{2\pi}{N} k (s-r)\right)
 \varphi_{s}\langle\varphi_{r},\cdot\rangle
\ .
\end{equation}
The sum over $k$ can be performed but it is advantageous to leave
it unperformed. We can now see that in the continuous limit, where
$N\rightarrow\infty$, $a\rightarrow 0$ and $g\rightarrow 0$ but
$agN\rightarrow $constant, the above commutator approaches the
value $i$, provided that lattice constants satisfy $agN=2\pi$. In
this limit, the sums over discrete indices $k, s, r,$  become
integrals over continuous variables $\kappa ,\sigma ,\rho ,$
according to the scheme
\begin{eqnarray}
 \sqrt{\frac{2\pi}{N}}k \rightarrow \kappa \ ,\
\sqrt{\frac{2\pi}{N}}s &\rightarrow& \sigma \ ,\
\sqrt{\frac{2\pi}{N}}r \rightarrow \rho \ ,\nonumber \\
\sqrt{\frac{2\pi}{N}}\varphi_{s} \rightarrow \varphi(\sigma)\ ,\
\sqrt{\frac{2\pi}{N}}\varphi_{r} &\rightarrow &\varphi(\rho)\ ,
\sum_{-j}^{j} \rightarrow \int^{\infty}_{-\infty} \ . \nonumber
\end{eqnarray}
The continuous limit is then given by
\begin{equation}
[X,P] \longrightarrow i\frac{agN}{2\pi } \int_{-\infty}^{\infty}
\!\! \!\!d\sigma \int_{-\infty}^{\infty} \!\! \!\!d\rho\
\frac{-1}{2\pi }\int_{-\infty}^{\infty} \!\! \!\!d\kappa\,\,
i\kappa (\sigma -\rho ) \exp\left(i\kappa(\sigma-\rho)\right)
 \varphi(\sigma)\langle\varphi(\rho),\cdot\rangle
\ .
\end{equation}
The sum over $k$, that was left unperformed, assumes in the
continuous limit a simple form. Indeed, the integral over $\kappa$
is a well known representation of Dirac's delta function.
Therefore
\begin{eqnarray}
[X,P] & \longrightarrow & i\frac{agN}{2\pi }
\int_{-\infty}^{\infty}\!\! \!\! d\sigma \int_{-\infty}^{\infty}
\!\! \!\!d\rho\,\, \delta (\sigma-\rho)
 \varphi(\sigma)\langle\varphi(\rho
),\cdot\rangle \nonumber \\ & = & i\frac{agN}{2\pi }
\int_{-\infty}^{\infty} \!\! \!\!d\sigma
 \varphi(\sigma)\langle\varphi(\sigma),\cdot\rangle = i\frac{agN}{2\pi }{\bf 1}
\ ,
\end{eqnarray}
where we have used the completeness relation. Therefore the usual
commutation relation for the continuous case is recovered,
provided that
\begin{equation}\label{agNeq}
 agN= 2\pi\ .
\end{equation}
\section{STATE AND TIME EVOLUTION}
At any instant of time, the state of the particle will be
determined by a Hilbert space element $\Psi$. We can represent
this state in the position or momentum representation, that is,
expanded in the bases $\{\varphi_{x}\}$ or $\{\phi_{p}\}$.
\begin{equation}\label{sta}
  \Psi=\sum_{x=-j}^{j}
  c_{x}\varphi_{x}=\sum_{p=-j}^{j}d_{p}\phi_{p}\ ,
\end{equation}
where the complex coefficients $c_{x}$ and $d_{p}$ are related by
the Discrete Fourier Transformation
\begin{equation}\label{dft}
d_{p} =\frac{1}{\sqrt{N}}\sum_{x=-j}^{j}
  \omega^{-px}c_{x}\ ,\
c_{x} =\frac{1}{\sqrt{N}}\sum_{p=-j}^{j}
  \omega^{px}d_{p}\ ,
\end{equation}
and their absolute values squared $|c_{x}|^{2}$ and $|d_{p}|^{2}$
represent the probability distributions for position and for
momentum. Let $\Psi(t_{0})$ be the state of the system at some
instant $t_{0}$, that we can choose to be $t_{0}=0$. In
Schr\"{o}dinger's picture, this state will evolve according to the
time evolution unitary operator given in terms of the hamiltonian
$H$ as
\begin{equation}
U_{t} = \exp(-iHt)\ .
\end{equation}
This description of the time evolution is equivalent to
Schr\"{o}dinger's equation if time is represented by a continuos
variable. However in some cases it may be convenient to assume
that also time takes discrete values giving preference to the
formulation with the time evolution operator above. If the state
is given in the position or in the momentum representation, the
coefficients of Eq.(\ref{sta}) will become functions of time
$c_{x}(t)$ and $d_{p}(t)$. Let us consider for instance the case
of a free particle with hamiltonian $H=P^{2}/2m$. In the momentum
representation the coefficients are simply given by
\begin{equation}\label{bdet}
 d_{p}(t)=d_{p}(0)\exp\left(-i\frac{g^{2}p^{2}}{2m}t\right)
 =d_{p}(0)\omega^{-p^{2}\frac{t}{\tau}}\ ,
\end{equation}
where we have introduced a \emph{time scale} $\tau$ defined by
\begin{equation}\label{tau}
 \tau=\frac{2ma}{g}\ ,
\end{equation}
whereas in the position representation we have
\begin{equation}\label{adet}
 c_{r}(t)=\sum_{x=-j}^{j}
  c_{x}(0)\frac{1}{N}\sum_{p=-j}^{j}
  \omega^{\left(p(r-x)-p^{2}\frac{t}{\tau}\right)}\ .
\end{equation}
The second summation in Eq.(\ref{adet}) is a Discrete Fourier
Transform that becomes, in the continuous limit, a Fourier
Integral Transform of a gaussian function with a very well known
result. In our discrete case, this summation can not be evaluated
in general. This is an example of the difficulties encountered in
the discrete case. It took Gau\ss\ four years working ``with all
efforts''\cite{gauss} in order to evaluate a similar summation
(the so called ``Gau\ss\ sum'') for some special values of the
parameters involved. In any case we can see that the state is
periodic, $\Psi(t+T)=\Psi(t)$ with period $T=N\tau$ if $N$ es odd
and $T=4N\tau$ if $N$ es even.
\section{STATE DETERMINATION  AND UNBIASED BASES}
At an early stage in the development of quantum mechanics, Pauli
\cite{paul} raised the question whether the knowledge of the
probability density functions for position and momentum where
sufficient in order to determine the state of a particle. Since
position and momentum are all the (classically) independent
variables of the system, it was, erroneously, guessed that this
Pauli problem could have an affirmative answer. Indeed, many
examples of Pauli partners, that is, different states with
identical probability distributions, where found. A review of
theses issues, with references to the original papers can be found
in refs.\cite{bal,wei1,wei2}. Considering the similar problem in
classical statistic, we should not be surprised to find out that
the Pauli question can not have a positive answer. The marginal
probability distribution functions of two random variables
uniquely determine the combined distribution function \emph{only} in the
case when they are uncorrelated, that is, when they are
independent random variables. Position and momentum are, however,
always correlated; that is indeed the essence of Heisenberg's
uncertainty principle, and therefore we should not expect that
their distributions uniquely determine the quantum state.
\par
Explicitly stated in our case, the Pauli question is: can we find
the set of $N$ complex numbers $\{c_{x}\}$ that determine the
state in  Eq.(\ref{sta}) with the knowledge of the sets
$\{|c_{x}|^{2}\}$ and $\{|d_{p}|^{2}\}$ related by Eq.(\ref{dft})?
Let us notice that the state has an arbitrary phase and is
normalized; therefore we only need to find $2N-2$ real numbers in
order to determine the state. The known numbers $\{|c_{x}|^{2}\}$
and $\{|d_{p}|^{2}\}$ are not independent because the numbers of
each set are probabilities and they should add to $1$. We have
therefore $2N-2$ equations at our disposal in order to find $2N-2$
unknown. However, the equations are \emph{not linear} and they are
not sufficient for an unambiguous determination of the state.
There is another very important feature in these equations. We
will see that not every set of data $\{|c_{x}|^{2}\}$ and
$\{|d_{p}|^{2}\}$ are compatible. The equations have solution only
if the position and momentum data satisfy a number of relations.
These constraint on the data is just Hiesenberg's uncertainty
principle and is a consequence of the relations in Eq.(\ref{dft}).
These concepts are clarified by an example with $N=2$.
\par
Let $\{\varphi_{-}, \varphi_{+}\}$ and $\{\phi_{-}, \phi_{+}\}$ be
the position and momentum bases in two dimensional Hilbert space.
Their internal product is given by Eq.(\ref{xp}). An arbitrary
state, normalized and with a phase fixed is determined by $2N-2=2$
numbers $0\leq\varrho\leq 1$ and $0\leq\alpha\leq 2\pi$:
\begin{equation}\label{n2sta}
 \psi=\varrho e^{i\alpha}\varphi_{-} +
 \sqrt{1-\varrho^{2}}\varphi_{+}\ .
\end{equation}
The independent data on position is
$|\langle\varphi_{-},\psi\rangle|^{2}=\varrho^{2}$, that directly
determines $\varrho$ and the independent data on momentum is
$|\langle\phi_{-},\psi\rangle|^{2}=\varpi^{2}$. With this last
data we must determine $\alpha$. Using that
$\langle\phi_{-},\varphi_{\pm}\rangle=\ \exp(\pm i\pi/4)/\sqrt{2}$, we
get after some algebra that
\begin{equation}\label{sina}
 \sin\alpha=\frac{\varpi^{2}-1/2}{\varrho\sqrt{1-\varrho^{2}}}\ .
\end{equation}
This equation can only have solution if
\begin{equation}\label{uncert}
 \left| \frac{\varpi^{2}-1/2}{\varrho\sqrt{1-\varrho^{2}}}\right|\leq
 1 \ ,
\end{equation}
that, after squaring and arranging, becomes
\begin{equation}\label{uncert1}
  (\varpi^{2}-1/2)^{2}+(\varrho^{2}-1/2)^{2}\leq (1/2)^{2} \ .
\end{equation}
 This relation is
indeed the uncertainty principle: if $\varrho^{2}=0$ or $1$, that
is, exact localization in $\varphi_{+}$ or $\varphi_{-}$, then
$\varpi^{2}=1/2$, that is, maximal spread in momentum and, vise
versa, exact momentum ($\varpi^{2}=0$ or $1$) implies maximal
spread in position ($\varrho^{2}=1/2$). Now, even if the data is
consistent with the uncertainty principle, there is an ambiguity
in the solution of Eq.(\ref{sina}) because if $\alpha$ is a
solution then $\pi-\alpha$ is also a solution. This ambiguity
\emph{can not} be solved with the given data and requires more
experimental information. We will next consider what observables
can we measure in order to determine the state without ambiguity.
\par
From previous example it is clear that we need further information
besides the distribution of position and of momentum in order to
determine the state of the particle. This will involve an
observable depending on both, position \emph{and} momentum because
any observable depending on only one of them will not bring new
independent information. Some candidates may be $X+P$ or the
correlation $XP+PX$ or any function $F(X,P)$ symmetric or
antisymmetric under the exchange $X\leftrightarrow P$. Perhaps the
best choice of an observable that provides information on the
system not available in the knowledge of position and momentum
distributions is an observable whose associated basis is
\emph{unbiased} to the position and to the momentum bases. Two
bases in a Hilbert space are unbiased when they are as different
as posible in the sense that any element of one basis has the same
``projection'' on \emph{all} elements of the other basis. More
precisely, the modulus of their internal product is a constant for
all pairs. Unbiased bases are candidates for the quantum
mechanical description of classically independent variables like
position and momentum; indeed we have from Eq.(\ref{xp})
$|\langle\varphi_{x},\phi_{p}\rangle|=1/\sqrt{N}\ \forall x,p$.
This leads us to the search of a basis $\{\eta_{s}\}$ unbiased to
$\{\varphi_{x}\}$ and to $\{\phi_{p}\}$.
\par
The importance of unbiased bases associated to non commuting
observables was recognized by Schwinger\cite{schwin} long ago but
the existence and explicit construction of maximal sets of
mutually unbiased bases for any dimension is still an open
problem. When the dimension $N$ is a prime number, a set of $N+1$
mutually unbiased bases was presented\cite{iva,woo} and this was
extended to the case when $N$ is a power of a prime
number\cite{woof}. In order to find unbiased bases we will follow
the method given by Bandyiopadhyay et al.\cite{band}. We have seen
that the eigenvectors $\{\phi_{p}\}$ of the operator $T$ that
produces a translation or \emph{shift} on the basis
$\{\varphi_{x}\}$ build an unbiased basis to $\{\varphi_{x}\}$.
This result is generalized in reference \onlinecite{band} where it
is shown that, if $N$ is prime, the eigenvectors of the unitary
operators $T, B, TB, TB^{2}, \cdots TB^{N-1}$ build a set of $N+1$
mutually unbiased bases where $T$ and $B$ are the translation
operators for position and momentum defined in Eqs.(\ref{defT}) and
 (\ref{defB}). In our case we want to find a set of only three unbiased
bases and therefore we just consider the first three operators
that provide unbiased bases for any $N$ (prime or not). The first
two operators provide the bases $\{\phi_{p}\}$ and
$\{\varphi_{x}\}$, that are related by the Discrete Fourier
Transform and are clearly unbiased. One can  easily prove, with
Eqs.(\ref{defT}, \ref{eigvB}) and (\ref{defB}, \ref{eigvT}) that
the operator $TB$ is a shift operator for \emph{both} bases
$\{\phi_{p}\}$ and $\{\varphi_{x}\}$ and therefore its
eigenvectors $\{\eta_{s}\}$ build a basis unbiased to both of
them. We have indeed
\begin{equation}\label{shift1}
  TB \varphi_{x}=\left\{\begin{array}{cc}
    \omega^{x}\varphi_{x+1} &\ , x\neq j \\
  \omega^{-2j^{2}}\varphi_{-j} &\ , x=j \
  \end{array}\right. \ ,
\end{equation}
and
\begin{equation}\label{shift2}
  TB \phi_{p}=\left\{\begin{array}{cc}
    \omega^{-(p+1)}\phi_{p+1} &\ , p\neq j \\
  \omega^{-2j^{2}}\phi_{-j} &\ , p=j \
  \end{array}\right. \ .
\end{equation}
 The eigenvectors of $TB$ in the position representation are
found by expanding $\eta_{s}$ in the basis $\{\varphi_{x}\}$ and
using Eqs.(\ref{shift1}) and the relation
\begin{equation}\label{eigvTB}
 TB \eta_{s} = \omega^{s} \eta_{s} \ .
\end{equation}
In this calculation one must use with care the modular mathematics
defined in Eq.(\ref{cycl}).  This results in
\begin{equation}\label{etapos}
\eta_{s}=\frac{1}{\sqrt{N}}\sum_{x=-j}^{j}
  \omega^{\frac{1}{2}x^{2}-(s+\frac{1}{2})x}\varphi_{x}
  \ .
\end{equation}
With a similar calculation we obtain the eigenvectors of $TB$ in
momentum representation
\begin{equation}\label{etamom}
\eta_{s}=\frac{1}{\sqrt{N}}\sum_{p=-j}^{j}
  \omega^{-\frac{1}{2}p^{2}-(s+\frac{1}{2})p}\phi_{p}
  \ .
\end{equation}
Clearly we see that $\{\eta_{s}\}$ is unbiased to
$\{\varphi_{x}\}$ and to $\{\phi_{p}\}$. The analytical
calculation of Discrete Fourier Transforms is much more difficult
than the Fourier Integral Transform and therefore it is of
mathematical interest that, from the last two equations we can
obtain the Discrete Fourier Transform for a family of sequences.
If we equate Eqs.(\ref{etapos}) and (\ref{etamom}) and we expand
 $\varphi_{x}$ in terms of $\{\phi_{p}\}$ we obtain
\begin{equation}\label{dft1}
 \frac{1}{\sqrt{N}}\sum^{j}_{x=-j}\omega^{\frac{1}{2}x^{2}-bx} \omega^{-px}=
 \omega^{-\frac{1}{2}p^{2}-bp}\ ,\left\{\begin{array}{ll}
    b=0,\pm 1,\pm 2, \cdots &\ , N\ \mbox{even}     \\
    b=\pm 1/2,\pm 3/2, \cdots   &\ , N\ \mbox{odd}
  \end{array}\right.\ .
\end{equation}
We rewrite this result in terms of the asymmetric indices more
usual in the mathematical literature as,
\begin{equation}\label{dft2}
 \frac{1}{\sqrt{N}}\sum^{N-1}_{n=0}(-1)^{n}\omega^{\frac{1}{2}n^{2}-bn}
  \omega^{-mn}= (-1)^{m}
 \omega^{-\frac{1}{2}m^{2}-bm}\ ,\forall b\ \mbox{integer}
 \ .
\end{equation}
The unbiased basis found is of course not unique. It was generated
by the operator $TB$ but there are many other operators whose
eigenvectors build unbiased basis to $\{\varphi_{x}\}$ and to
$\{\phi_{p}\}$. Indeed, any combination $T^{n}B^{m}$ or
$B^{m}T^{n}$ where $n$ and $m$ are not divisors of $N$ could do
the job.
\par
We will now try to find the physical meaning of the basis
$\{\eta_{s}\}$. That is, we would like to find an hermitian
operator $S(X,P)$ with $\{\eta_{s}\}$ as eigenvectors
corresponding to the eigenvalues $hs$ with $h=ag$. That is,
\begin{equation}\label{sxp}
 TB=\exp\left(iS(X,P)\right)\ .
\end{equation}
In terms of the operators $X$ and $P$, and using the relation
$agN=2\pi$, we have
\begin{equation}\label{sxp1}
  \exp\left(iS(X,P)\right)=
  \exp\left(-iaP\right)\exp\left(igX\right)\ .
\end{equation}
Notice that here we \emph{can not} use the
Baker-Campbell-Hausdorff relation $e^{P}e^{X}=e^{P+X-i/2}$ that is
valid in the $N\rightarrow\infty$ case, where the commutator
$[X,P]$ is a constant. If this were possible, then the operator
$S$ would be simply equal to $gX-aP$. This is one of the subtle
differences of finite and infinite dimensional Hilbert spaces. It
is posible to find the eigenvectors of the operator $X-P$ but the
basis so obtained is \emph{not} unbiased to either
$\{\varphi_{x}\}$ nor $\{\phi_{p}\}$ however it becomes unbiased
in the infinite dimensional limit\cite{dlt}. The relation of $S$
to $X$ and $P$ is not simple but we can prove that $S$ is
antisymmetric under the exchange $P\leftrightarrow X$ and
$a\leftrightarrow g$. Indeed, from the hermitian conjugate of
Eq.(\ref{sxp1}) we have
\begin{equation}\label{sxp2}
  \exp\left(-iS(gX,aP)\right)= \exp\left(-igX\right)\exp\left(iaP\right)
  = \exp\left(iS(aP,gX)\right)\ .
\end{equation}
\section{CONCLUSION}
In this work we have presented the quantum mechanical formalism
for position and momentum of a particle in a one dimensional
cyclic lattice in a way that may be useful for a didactic
complement of the infinite dimensional case presented in quantum
mechanics text books. In doing this, several mathematical
subtleties related to the difference between infinite and finite
dimensional Hilbert spaces, and of modular mathematics, arise. We
have discussed the physical and mathematical relevance of unbiased
bases and, as consequence from the construction of such a basis,
the Discrete Fourier Transform for a family of sequences is given.
\par
It is a strongly recommended exercise to reproduce all this work
in terms of the asymmetric indices running from $0$ to $N-1$. One
can see thereby the need to define the translation operator always
cyclic in order to get the position and momentum bases related by
the Discrete Fourier Transform. The calculations of the
eigenvectors of the operator $TB$ are useful exercises for the
modular mathematics.
\par
This work received partial support from ``Consejo Nacional de
Investigaciones Cient\'{\i}ficas y T\'ecnicas'' (CONICET),
Argentina.
\section{APPENDIX A}
All sums appearing in this appendix can be derived from the
fundamental expression
\begin{equation}\label{sum0}
  \sum^{N-1}_{k=0}z^{k}=\frac{1-z^{N}}{1-z} \ ,
\end{equation}
 for any  complex number $z$, that in terms of the symmetric index becomes
 \begin{equation}\label{sum1}
  \sum^{j}_{k=-j}z^{k}=\frac{z^{N/2}-z^{-N/2}}{z^{1/2}-z^{-1/2}}
  \mbox{ for }
  \left\{\begin{array}{ll}
    j=\frac{1}{2}, 1, \frac{3}{2}, 2, \cdots \\
   N=2j+1=2, 3, 4, 5, \cdots  \\
   z \mbox{ complex}
  \end{array}\right.\ .
\end{equation}
If $z$ takes the special values
$z=\exp(i\frac{2\pi}{N}r)=\omega^{r}$ with $r$ an arbitrary
number, then
\begin{equation}\label{sum2}
  \sum^{j}_{k=-j}\omega^{kr}=\frac{\sin (\pi r)}{\sin (\frac{\pi}{N} r)}
\ .
\end{equation}
In our case, the number $r$ will often assume integer or
half-odd-integer values. For these cases we have,
\begin{equation}\label{sum3}
  \sum^{j}_{k=-j}\omega^{kr}=
\ \left\{\begin{array}{ll}
 (-1)^{n(N-1)}N   &\mbox{ for } r=nN,\ n=0,\pm 1,\pm 2, \cdots \\
 0  &\mbox{ for } r= \pm 1,\pm 2, \pm 3, \cdots  \neq nN \\
\frac{2\omega^{\frac{r}{2}}}{1-\omega^{r}} &\mbox{ for } r= \pm
\frac{1}{2},\pm \frac{3}{2}, \pm \frac{5}{2}, \cdots
  \end{array}\right.\ .
\end{equation}
The first two cases correspond, with the asymmetric index, to the
well known result $ \sum^{N-1}_{k=0}\omega^{kr}= N\delta_{r,nN}$
where we see that this choice leads to simpler mathematics. The
third case above must be handled with care in numerical
evaluations because the numerator is the \emph{fourth root} of
$\exp(i\frac{2\pi}{N}u)$, with $u$ an odd integer. Therefore it
has four posible numerical results. The denominator has also two
posible results. A formal derivative of Eq.(\ref{sum2}) with
respect to the parameter $r$ leads to the result
\begin{equation}\label{sum4}
\sum^{j}_{k=-j}k\omega^{kr}=\frac{i}{2}\frac
 {\sin (\pi r)\cos (\frac{\pi}{N} r)-N\cos (\pi r)\sin (\frac{\pi}{N} r)}
{\sin^{2} (\frac{\pi}{N} r)} \ .
\end{equation}
Deriving again with respect to $r$, we can obtain other summations
involving higher powers of $k$.
\section{APPENDIX B}
In most textbooks it is proven that the non degenerate eigenvalues
of an hermitian operator are real and their eigenvectors are
orthogonal. We give here the corresponding proof for
\emph{unitary} operators.
\par
 Let $T$ be an unitary operator and $\lambda_{k}$ and $\phi_{k}$
 the non degenerate
eigenvalues and normalized eigenvectors. Then, we will prove that,
$|\lambda_{k}|^{2}=1$ and
$\langle\phi_{r},\phi_{k}\rangle=\delta_{rk}$.
\\ From $T\phi_{k}=\lambda_{k}\phi_{k}$ and $T^{\dag}T=1$ it
follows that
\begin{equation}
|\lambda_{k}|^{2}= \langle T\phi_{k},T\phi_{k}\rangle = \langle
\phi_{k},T^{\dag}T\phi_{k}\rangle = 1\ .
\nonumber
\end{equation}
In order to prove the orthogonality consider that
\begin{eqnarray}
 T\phi_{k}=\lambda_{k}\phi_{k} &\rightarrow&
\langle\phi_{r},
T\phi_{k}\rangle=\lambda_{k}\langle\phi_{r},\phi_{k}\rangle\ ,
 \nonumber \\
 T^{\dag}\phi_{r}=\lambda^{*}_{r}\phi_{r} &\rightarrow &
\langle T^{\dag}\phi_{r},
\phi_{k}\rangle=\lambda_{r}\langle\phi_{r},\phi_{k}\rangle \ .
\nonumber
\end{eqnarray}
Subtracting both equations we get
$0=(\lambda_{k}-\lambda_{r})\langle\phi_{r},\phi_{k}\rangle$.
Since the eigenvalues are non degenerate, the product
$\langle\phi_{r},\phi_{k}\rangle$ must vanish for $k\neq r$.
\par
Since $T$ is bounded, the completeness of the eigenvectors can be
proved in the usual way and therefore $\{\phi_{k}\}$ is a basis.

\end{document}